# High sensitivity and large scanning range optical antennas enabled by multi-casting ridge-waveguide subwavelength structure arrays

**WEIJIE XU,[1] XIANXIAN JIANG,[1] YELONG BAO,[1] AND JUNJIA WANG[1,*]**

[1]*National Research Center for Optical Sensors/Communications Integrated Networks, School of Electronic Science and Engineering, Southeast University, Nanjing, 210096, China*
*\*junjia_wang@seu.edu.cn*

**Abstract:** With the rapid development of large-scale integrated photonics, optical phased array (OPA) is an effective way to realize highly integrated, stable and low-cost beam control system. Achieving a large field of view (FOV) in the longitudinal direction without increasing fabrication cost and system complexity is still a significant challenge in OPA antennas. Here, a high sensitivity and large scanning range antenna based on subwavelength structure array is proposed to enhance the longitudinal scanning and free-space radiating efficiency by using the ridge-waveguide structure and backward-emitting. A millimeter-long grating antenna with a far-field beam divergence of 0.13° and a wavelength sensitivity of 0.237°/nm is experimentally demonstrated. Furthermore, by using different sideband periods, we introduce a multi-casting grating antenna with a large scanning range up to 42.6°. The proposed devices show significant improvement in longitudinal wavelength sensitivity compared with the typical waveguide grating antennas.

## 1. Introduction

Optical phased array (OPA) is a low-cost[1-3], large-scale[4-6] nanophotonic system capable of generating and manipulating optical beams, which typically includes laser sources, beam splitters, phase shifters, and emitters[7-11]. High-performance OPA exhibits outstanding real-time accurate shaping and scanning for low-loss[12-14], large field of view (FOV)[15-17], and high-speed two-dimensional (2D) beam steering, which provides a new solution for important fields such as light detection and ranging (LiDAR), free-space data communication, and optical imaging[18-23].

Integrated optical antennas are one of the key components of OPAs[24-26], which convert transmitted light into radiation power in free space, providing a nanoscale compact, lightweight footprint for large-scale arrayed integration[27-32] without relying on complex mechanical transceiver components[33-38]. The beam divergence angle is an important parameter influencing OPA, which is determined by the length of the optical antenna[5, 39, 40]. Various material platforms have been explored to overcome this limitation[14, 41-46]. Due to the relatively low refractive index contrast on the silicon nitride ($Si_3N_4$) and the ultra-low optical propagation loss on the LN (lithium niobate) platform, millimeter-scale optical antenna length can be easily achieved, which consequently generates a narrow beam divergence. But these methods are limited in beam radiation efficiency, while multilayer structures and different material platforms introduce manufacturing complexities, which are unsuitable for large-scale, ultra-compact integration. As an important platform for photonic integration, silicon on insulator (SOI) is considered as a promising option for achieving high-precision, robust, and low-cost beam control systems[47, 48]. Whereas the high refractive index contrast between silicon and silica leads to strong mode confinement. There have been many efforts to extend the length of grating antenna, such as etching the surface to form shallow etched grating[29, 49] and sidewall corrugation to form corrugated waveguide[50, 51], which can adjust the relative refractive index within a certain range but still remain strong grating strength. In addition, sidewall corrugation increases the crosstalk in adjacent waveguide channels, resulting in undesired power consumption. Hence, further exploration in SOI is required to reduce the grating strength for achieving narrow beam divergence in the far-field.

Subwavelength structures have emerged as one of the effective ways to alter the macroscopic optical properties of materials in integrated photonics[52-57]. Owing to the flexibility in adjusting the effective refractive index, it is widely used in next-generation optical communications[58-60], biomedical[61, 62], quantum[63, 64] and sensing technologies[65, 66]. Many fundamental components based on subwavelength structure for high-performance integrated optical systems are demonstrated, including beam splitting[67-69], polarization[70-73], spectral filtering[74-77], and couplers[78-80]. Furthermore, there have been studies on subwavelength antennas, where weak grating strengths are achieved by placing subwavelength segments on both side of the central waveguide to balance the small beam divergence angle and large FOV[81-83].

The steering range is another figure of merit affecting the FOV of an OPA. However, the longitudinal wavelength sensitivity of most grating antennas is 0.14-0.17°/nm, which leads to a narrow longitudinal scanning range. The slow light in photonic crystal waveguides was used to realize the large group index for improving longitudinal wavelength sensitivity[84-86]. But the high loss of photonic crystal structure is difficult to enable long gratings. A large longitudinal scanning range is realized at the system level by utilizing polarization switching[87-89]. Nevertheless, the implementation of polarization controlling increases the overall power consumption and structural complexity. In general, grating antennas are usually operated in forward-emitting mode, and the potential advantages of backward-emitting mode for wavelength sensitivity have rarely been discussed[81, 90]. Besides, the utilization of multi-casting to achieve multi-beam scanning may be an prospect to increase the longitudinal scanning range, but most of the relative concepts are used in the microwave range[91-93].

In this paper, we proposed grating antennas to enhance the longitudinal scanning and free-space radiating efficiency by introducing multi-casting ridge-waveguide subwavelength structure arrays. The backward-emitting approach is used to improve the wavelength tuning performance in the longitudinal direction without adding structural complexity. We demonstrate a ridge-waveguide subwavelength antenna which is able to achieve a longitudinal wavelength sensitivity of 0.237°/nm, with a narrow beam divergence of 0.13°. Moreover, a multi-casting subwavelength antenna is developed, with double-beam emission in the far-field by setting different grating periods on both sidebands. A large scanning range of 42.6° is achieved. The proposed subwavelength antennas provide a promising solution for integrated OPAs with high sensitivity and large scanning range.

## 2. Device configuration

Our device, illustrated in Fig. 1(a), consists of a central ridge-waveguide and an array of ridge-waveguide subwavelength structures symmetrically placed on both sides. A double-layer taper with $L_t$=10 μm is used to reduce the transmission loss between the interconnecting waveguide and the antenna. We consider the condition where two sidebands with same grating period ($\Lambda_1 = \Lambda_2$). The proposed antenna is simulated by 3D finite-difference time-domain (FDTD) with a silicon thickness $H_1 = 220$ nm and a buried oxide (BOX) layer thickness of 3 μm. Then, we consider grating strength as the figure of merit (FOM). The particle swarm algorithm (PSA) is used to optimize the parameters ($\Lambda_1, \Lambda_2, L, W_3, W_4$) and spacing $d$ of the ridge-waveguide subwavelength structure arrays. Table 1 summarized the optimized parameters of the grating antenna. A weak grating strength of 1.2 mm$^{-1}$ is achieved with millimeter-long emission.

**Table 1. Standard structural dimensions of the proposed grating antenna (µm)**

| $H_1$ | $H_2$ | $W_1$ | $W_2$ | $W_3$ | $W_4$ | L | $L_t$ | d | $\Lambda_1$ | $\Lambda_2$ |
|---|---|---|---|---|---|---|---|---|---|---|
| 0.22 | 0.07 | 0.4 | 0.11 | 0.32 | 0.11 | 0.195 | 10 | 0.23 | 0.65 | 0.65 |

Subwavelength sidebands enable flexible control of the grating strength. The transverse-electric (TE) field intensity distribution of the fundamental modes in antenna without and with subwavelength sidebands are shown as insets (1) and (2) in Fig. 1(b). The rectangular blocks with $H_2$ = 70 nm on top of the center strip waveguide and subwavelength structure array are added to form the ridge-waveguide structure. It can be seen that the electric field is enhanced at the ridge-waveguide corners. The ridge-waveguide structures are used to enhance the longitudinal diffraction ability and sensitivity of the propagation light while maintaining the weak grating strength, as will be detailed in the next section.

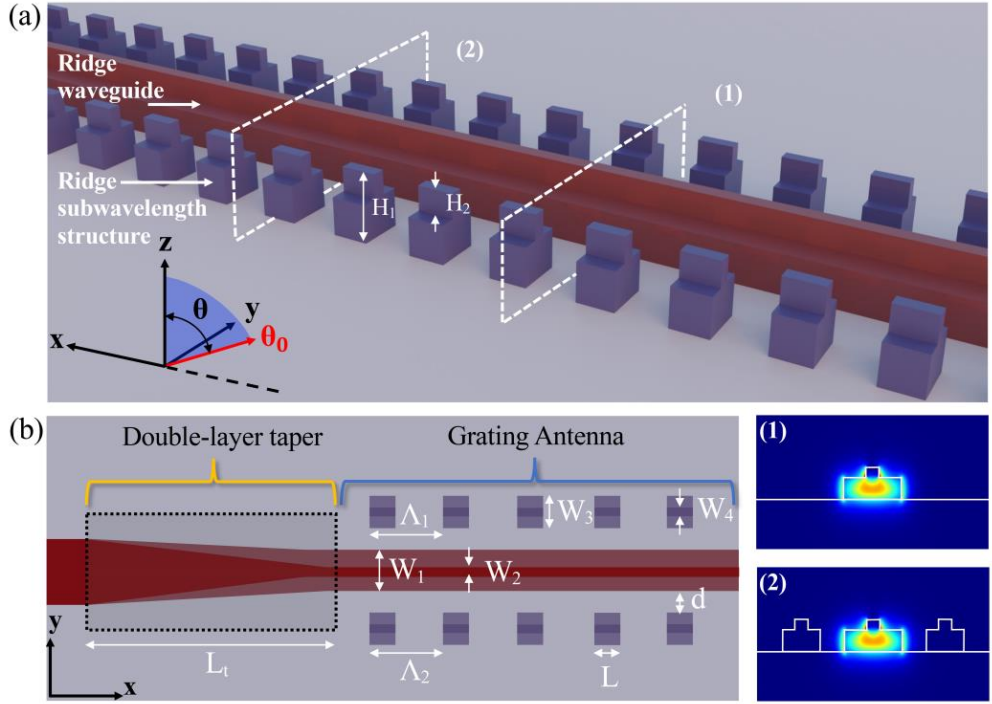

**Fig. 1.** (a) 3D schematic of the proposed antenna. Inset: Electric field distribution of the fundamental mode supported by section without (1) and with (2) ridge-waveguide subwavelength block. For visual clarity, different color is added with the Ridge-waveguide shaded as red, and the Ridge-waveguide subwavelength structure shaded as purple. (b) Top view of the input part of the antenna, including input waveguide with double-layer taper and grating antenna.

## 3. Simulation results

The grating strength of the antenna can be flexibly adjusted by modifying the dimensions of the structure, which can be expressed as $\alpha=\ln(P_{out}/P_{in})/(-2L_{Antenna})$. We simulate the relation between grating strength and structural dimensions at a wavelength of 1550 nm. The period of the grating $\Lambda_1(\Lambda_2)$ is selected as 0.65 µm to ensure that the antenna is in backward-emitting mode. The results are summarized in Fig. 2, where the parameters we used are highlighted with red crosses. Fig. 2(a) and (b) show that the grating strength can be reduced by increasing $W_1$ or decreasing $W_3$. However, it is less sensitive to $W_2$ and $W_4$. Since they are used to improve the radiation efficiency. Fig.2(c), (d) and (e) demonstrate the relation of d with $W_3$, $W_4$ and $FF$, respectively. *FF* is the filling factor of the grating, which is defined as *FF* = L/Λ. It can be seen that weak grating strengths can be easily achieved by changing *d*. The white dashed line in Fig.2 (f) indicates *d* = 0.23 μm. It can be observed that the grating strength remains practically constant within the working wavelength band (1500-1600 nm), which means that far-field beam divergence is also consistant with wavelength, according to the relationship between the divergence angle and the length of the antenna. Moreover, the spot distribution in the far-field is also a crucial factor of the antenna. Therefore, the structural dimensions shown in Table 1 are obtained by comprehensive evaluation.

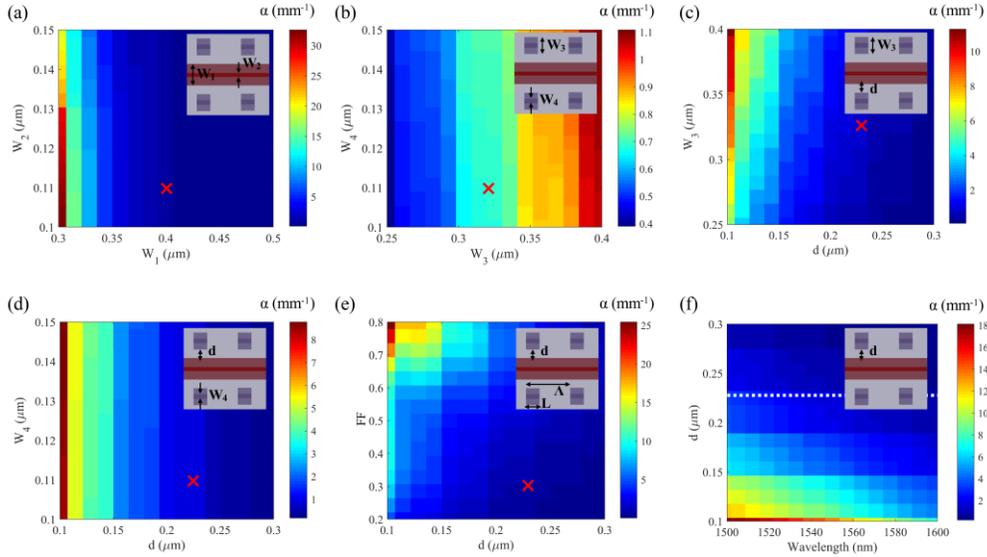

**Fig. 2.** Grating strength variation in relation to (a) $W_1$ and $W_2$, (b) $W_3$ and $W_4$, (c) *d* and $W_3$, (d) *d* and $W_4$, (e) *d* and *FF*, (f) wavelength and *d*. The parameters we used are highlighted with red crosses, and the white dashed line indicates *d* = 0.23 μm.

The beam emitted by the proposed grating antenna in the far-field region is divergent in the lateral direction ($\phi$), and collimated in the $\theta$ direction. As indicated in Fig. 1(a), $\theta_0$ represents the grating antenna emission angle. The grating equation can be used to design beam steering by tuning wavelength. According to the Bragg condition[33]:

$$\sin(\theta_{air}) = n_{clad} \sin(\theta_{clad}) = n_{eff}(\lambda) - \frac{\lambda}{\Lambda} \qquad (1)$$

where $\theta_{air}$ and $\theta_{clad}$ are the emission angles in air and upper cladding, respectively. $n_{clad}$ is the refractive index of the cladding (here is air), $n_{eff}$ is the effective refractive index of the guiding mode, and Λ is the period of the grating.

A K-space diagram was utilized to demonstrate the light diffraction induced by the grating antenna to explore the wavelength tuning ability of the antenna in the $\theta$ direction, as shown in Fig. 3(a). The grating antenna is assumed to propagate along the x-axis with grating vector GK=m·2π/Λ. It diffracts into the air when the guided mode propagates in the grating with propagation constant $\beta = (2\pi \cdot n_{eff})/\lambda$. Emission angle ($\theta$) is determined by the phase-matching condition between the propagation constant $\beta$ and the grating vector GK. Here we define the grating antennas at two different wavelengths $\lambda_1$ and $\lambda_2$, corresponding to diffraction angles $\theta_1$ and $\theta_2$ in air. according to the grating Eq. (1):

$$\sin(\theta_1) - \sin(\theta_2) = n_{eff}(\lambda_1) - n_{eff}(\lambda_2) + \frac{\lambda_2}{\Lambda} - \frac{\lambda_1}{\Lambda} = \Delta n_{eff}(\lambda) + \frac{\Delta\lambda}{\Lambda} \qquad (2)$$

And we find $|\sin(\theta_1) - \sin(\theta_2)| \leq |\theta_1-\theta_2|$, according to Lagrange's median theorem. Moreover, the equation can be expressed as:

$$\Delta\theta \geq \Delta n_{eff}(\lambda) + \frac{\Delta\lambda}{\Lambda} \qquad (3)$$

Assuming that $\lambda_1$=1500 nm, $\lambda_2$=1600 nm, it can be obtained that $n_{eff}(\lambda_1)>n_{eff}(\lambda_2)$ according to the simulation result of $n_{eff}$ in Fig. 3(d), i.e., $\Delta n_{eff}(\lambda)>0$. Therefore, Eq. (3) can be rewritten as:

$$\Delta\theta \geq \frac{\Delta\lambda}{\Lambda} \qquad (4)$$

for a wavelength tuning range of 100 nm, Eq. (4) can be simplified to $|\Delta\theta/\Delta\lambda| \geq 0.1/\Lambda$, where $|\Delta\theta/\Delta\lambda|$ is the wavelength sensitivity of the grating antenna. It can be seen that the wavelength sensitivity is inversely proportional to the grating period.

Then, we scan the period of the proposed grating antenna to obtain the wavelength sensitivities at different periods, simulated by 3D FDTD, as shown in Fig. 3(b). It can be seen that the simulation results are in good agreement with the theoretical prediction. By reducing the grating period to less than 0.95 μm, the grating antenna switch to backward-emitting mode, and enhances beam scanning ability in the $\theta$ direction simultaneously. However, keep reducing the grating period causes an unexpected change in the diffraction angle at certain wavelengths, where the radiation loss of the antenna increases dramatically. Since it crosses the Bragg reflection regime.

The ridge-waveguide structure is introduced to the grating antenna by using the blazing effect[94] to break the vertical symmetry of the space, which enhances the diffraction ability in the longitudinal direction ($\theta$). Fig. 3(c) demonstrates that the ridge-waveguide structure can increase the radiation efficiency of the grating antenna, with an average value of 70%. In addition, it can be calculated from Fig. 3(d) that $\Delta n_{eff}(\lambda)|_{with} > \Delta n_{eff}(\lambda)|_{without}$ when the wavelength ranges from 1500 nm to 1600 nm. Therefore, the ridge-waveguide structure can also improve the wavelength sensitivity according to Eq. (3). Fig. 3(e) shows the wavelength sensitivity of the grating antenna without ridge-waveguide structure is 0.192°/nm, while it increases to 0.242°/nm with the ridge-waveguide structure, which leads to a significant improvement of scanning range.

Fig. 3(f) shows that our simulation predicts the beam radiating angle from $\theta$ = -37.7° to $\theta$ =-42.5° when the operational wavelength is scanned from 1530 to 1570 nm with 10 nm spacing. The corresponding far-field distribution in the $\theta$ direction is plotted in the color inset of Fig. 3(f). It shows that the proposed grating antenna can realize single spot in the far-field and the radiation patterns are well confined without interference pattern.

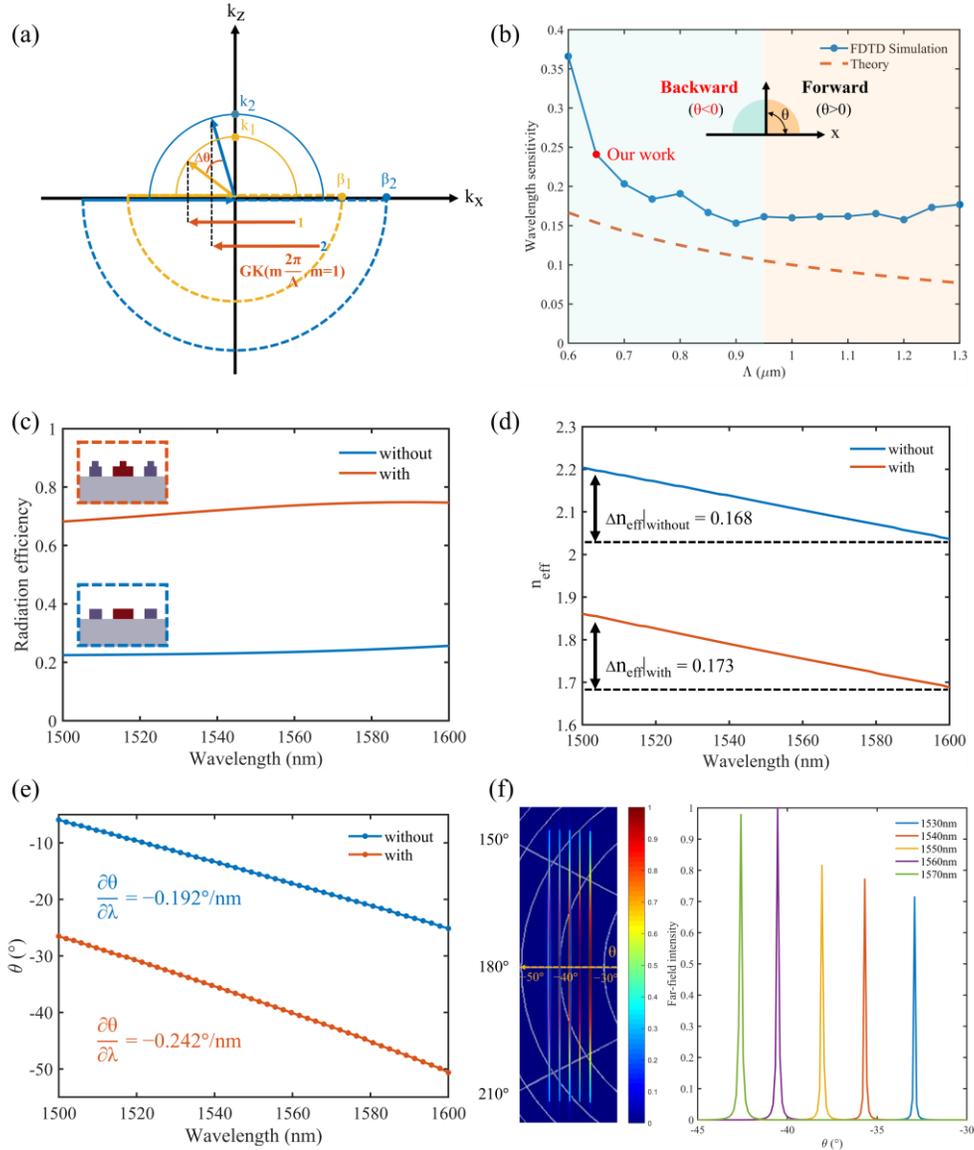

**Fig. 3.** (a) K-space diagrams of grating antennas corresponding to two different wavelengths. (b) Wavelength sensitivity as a function of the grating period. (c) Radiation efficiency of the grating antennas without and with ridge-waveguide structures. (d) Effective refractive index of the guiding mode without and with ridge-waveguide structures. (e) Wavelength sensitivity of the grating antennas without and with ridge-waveguide structures. (f) The far-field profile along θ direction of the proposed antenna when wavelength sweeps from 1530 nm to 1570 nm, the color inset is the corresponding far-field distribution.

Furthermore, we can exploit the different grating periods ($\Lambda_1 \neq \Lambda_2$) of subwavelength sideband antennas with double-beam emissions to expand our scanning range. First, the grating period is designed according to the grating equations to achieve a wider scanning range. Fig. 4(a) shows two grating antennas with different sideband periods. Subwavelength sideband 1 and 2 are capable of achieving beam steering of 18.4° and 23.2° in the wavelength range from 1500 nm to 1600 nm, respectively. Then, combine the two sidebands realize a multi-casting ridge-waveguide grating antenna, as shown in Fig. 4(c).

The emission angle of the antenna in $\theta$ direction as a function of wavelength is shown in Fig. 4(b). It can be observed that the grating antenna 3 enables double-beam emission when the operating wavelength is scanned from 1500 nm to 1600 nm, where the beam angle of sideband 1 at 1600 nm exactly corresponds to the sideband 2 at 1500 nm, as shown by the purple line. Thus, a beam steering angle of 42.6° in the far-field is realized, as shown in the yellow region. Compared to the conventional configuration, multi-casting antenna have power loss in each beam because it emits several beams at different angles. The simulated power loss value is 3.9dB for our case. Nevertheless, this trade-off is worthwhile for enlarging the scanning range. Notably, the design concept can be further used to configure multi-channel or cascaded arrays with different sideband periods to realize multi-beam scanning in the far-field, which provides a potential candidate for the implementation of ultra-wide scanning optical phased arrays.

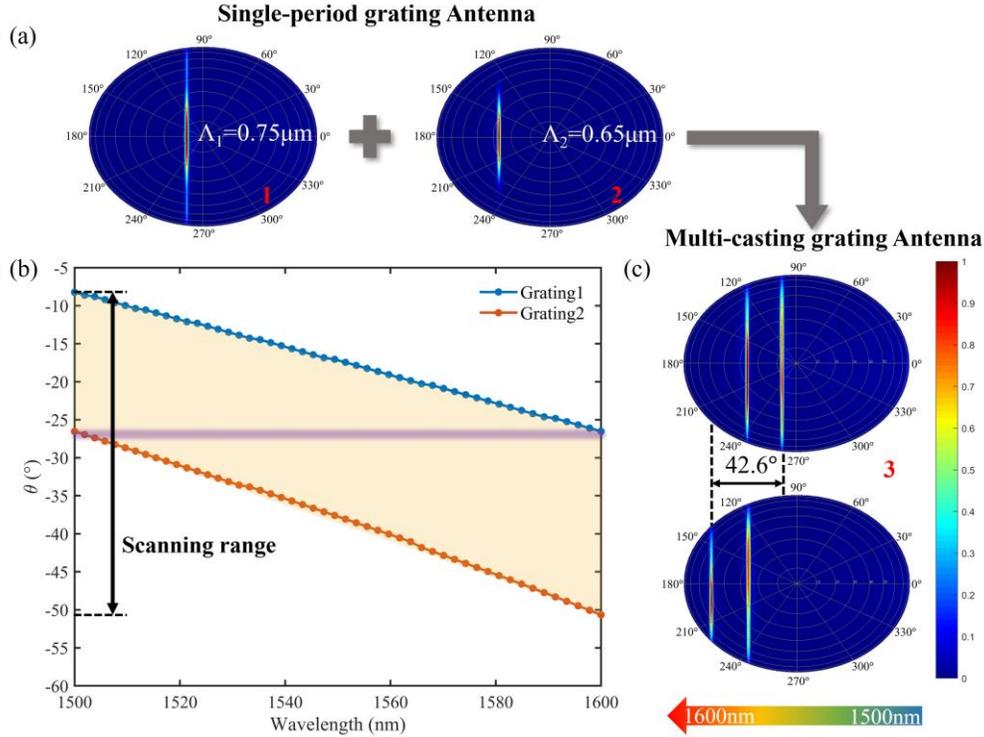

**Fig. 4**. (a) Far-field wavelength tuning distributions for grating antenna 1 and 2 with periods $\Lambda_1 = 0.65$ μm and $\Lambda_2 = 0.75$ μm, respectively. (b) Emission angle as a function of wavelength for multi-casting ridge-waveguide grating antenna (c) Far-field wavelength tuning distribution of proposed multi-casting grating antenna 3. Note that the wavelength tuning range for all far-field distributions is 100 nm.

## 4. Experimental Results

The device was then fabricated on a 220 nm SOI platform with a 3-μm-thick BOX layer. The chip was patterned using electron beam lithography (EBL) and a two-step reactive ion etching (RIE) processes ($H_1$ = 220 nm deep etch and $H_2$ = 70 nm shallow etch). The leakage loss from the silicon substrate is negligible[95]. Fig. 5(c) shows scanning electron microscope (SEM) image of a section of the fabricated grating antenna, where a clear ridge-waveguide and subwavelength structures can be seen.

We use an Agilent 8164B tunable laser to scan the wavelength of the light and a polarization controller is used to select TE polarization at the chip input. The light was coupled into the chip by the grating coupler and a double-layer taper was used to connect the optical antenna under test. We measure the transmission loss by connecting a grating coupler and we remove the grating coupler to measure the far-field characteristics. An optical power meter is used to analyze the output light for measuring grating strength. Then, we measure the scanning range and wavelength sensitivity ($\partial\theta/\partial\lambda$) of the antenna by an infrared camera (Xenics Bobcat 320) with 320×256 pixels at 20um pitch and an ATW SWIR infrared shortwave lens with a focal length $f$ = 35 mm. The overall test system is shown in Fig. 5(a). The angular resolution of the single lens setup can be estimated by $\Delta\theta=\tan^{-1}(\Delta\text{pixel}/f)$ according to the geometrical relationship, where $\Delta$pixel and $f$ represent the pixel size of the IR camera and the focal length of the imaging lens, respectively. The FOV can be estimated by $\theta_{th}(\phi_{th})=\tan^{-1}(S/f)$, where S represents the size of the receiving area of the IR camera. Thus, $\Delta\theta$ and FOV are estimated to be 0.033° and 10.4°×8.3°, respectively, and such a configuration is sufficient to test the designed grating antennas. The schematic of the setup is shown in Fig. 5(b). The infrared InGaAs sensor array is located at the focal plane of the lens, and the imaging system colletcts the far-field distribution of the antenna by using the 2D Fourier transform of the near field[96]. Backward-emitting mode represents the steering angle θ<0, and the position of the IR camera is set to capture the beam emitted by the proposed grating antenna.

We measured the grating strength of the optical antenna first. Reference waveguide is used to exclude the influences of transmission loss and coupling loss. As shown in Fig. 5(d), the red line is the theoretical value calculated according to Transmission(dB)=$-20\alpha L_{Antenna}\cdot\log_{10}e$, where $\alpha$ and $L_{Antenna}$ represent the grating strength and the length of the antenna, respectively. The optical antenna is set to different lengths, 585 μm, 780 μm, and 975 μm, and the blue circles are the measured transmission loss of the corresponding lengths, which are in excellent agreement with the simulation, thus obtaining the grating strength $\alpha$ = 1.2 mm$^{-1}$.

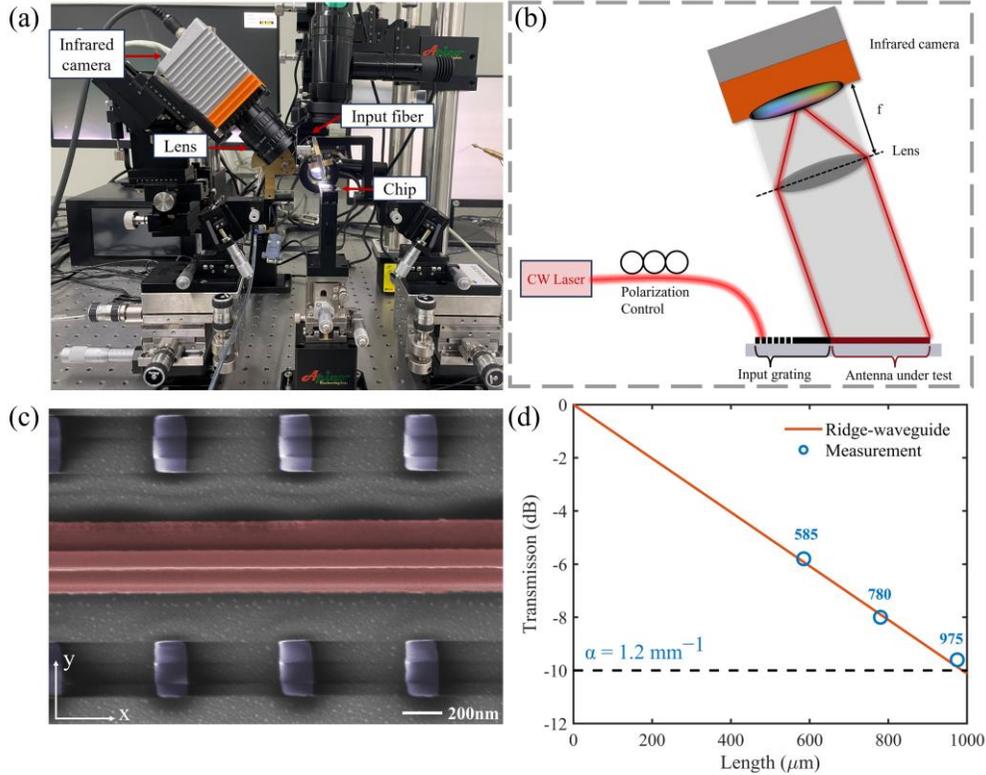

**Fig. 5.** (a) The test system setup. (b) Schematic of the imaging system. (c) SEM image of the grating antenna, with $\Lambda_1 = \Lambda_2$. (d) Transmission loss of the proposed antenna as a function of length.

Then, we measured the far-field characteristics of the proposed antenna. The operating wavelength is scanned from 1540 nm to 1565 nm to capture the beam steering range in the θ direction. The selection of a small wavelength range is attributed to the limitation of the infrared camera's FOV. Fig. 6(a) shows the far-field image collected by the IR camera. As expected, the beam of the optical antenna is collimated in the θ direction and diverged in the $\phi$ direction. Note that due to the setup of the test system, the middle section of the far-field beam emitted in the backward direction is blocked by the input fiber, resulting a discontinuity in the middle of the measurement. Fig. 6(b) shows the measured emission angle of the antenna in the $\theta$ direction. It can be observed that the beam angle changes from -35° to -40.93° when sweeping wavelength from 1540 nm to 1565 nm. The wavelength sensitivity calculated from the slope is 0.237°/nm, which agrees well with the simulation result of 0.241°/nm. The measured far-field beams corresponding to wavelengths along the θ-direction are shown in Fig. 6(c), where the intensity of the far-field beam varies with wavelength due to the coupling loss of the input grating coupler. The divergence angle distribution of the optical antenna at 1550 nm is shown in the inset of Fig. 6(b), which is measured to be 0.13°. The difference in the divergence angle of the beam can be attributed to the variation in grating strength caused by process manufacturing deviations and phase errors due to manufacturing defects, as mentioned in other far-field experiments[11, 41, 81, 83].

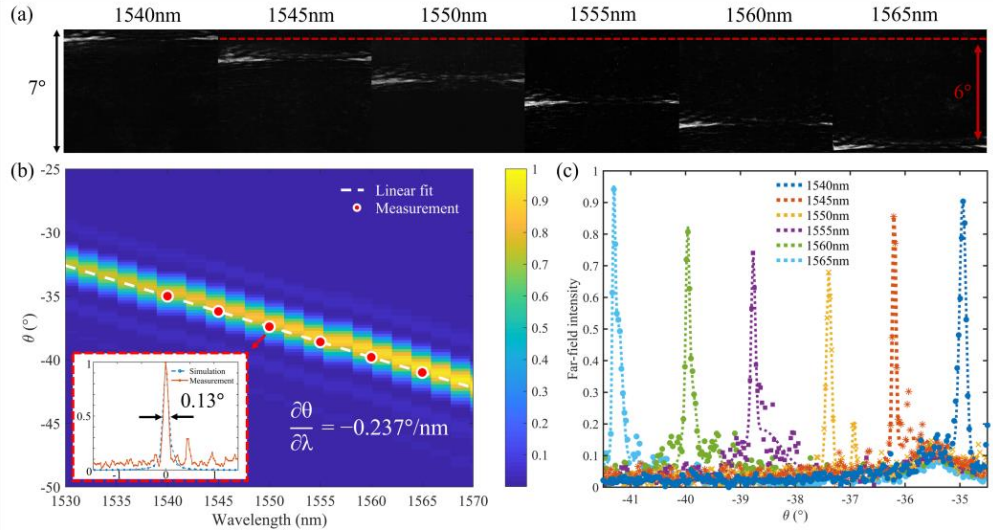

**Fig. 6.** (a) Far-field images measured by infrared cameras for wavelengths range from 1540 to 1565 nm. (b) The measured emission angle as a function of wavelength. The heat map is the corresponding simulated far-field normalized intensity, and the wavelength sensitivity of the antenna is estimated as the slope of the linear fit. The inset shows the measured beam width at 1550 nm along the θ direction. (c) Measured far-field beam intensity along θ direction for the corresponding wavelengths.

## 5. Discussion and Conclusion

Table 2 summarizes the various performance metrics of the current state-of-the-art waveguide grating antennas. In contrast, to achieve similar emission performance, either extremely small feature sizes or other material platforms such as SiN or LN (lithium niobate) are required, which undoubtedly increase the complexity and difficulty of the fabrication process. The proposed solution in this paper shows that weak large-scale optical antennas can be achieved in the SOI platform while maintaining an ultra-high wavelength sensitivity.

In summary, we experimentally demonstrate a high sensitivity and large scanning range optical antenna enabled by ridge-waveguide subwavelength structures. The proposed antenna exhibits a far-field beam divergence angle of 0.13° in the θ-direction, with a wavelength sensitivity of 0.237°/nm and a minimum feature size of 110 nm. It can be easily fabricated by photolithography, which reduce the fabrication cost and system complexity. The proposed device is capable of multi-casting, where a dual-beam emission in the far-field by introducing subwavelength sidebands with different periods, thus realizing a large scanning range of 42.6° in the $\theta$ direction. It can also be further configured as a multi-channel, large-aperture OPA, demonstrating great potential for future free-space interconnect and LIDAR applications.

**Table 2. State of art waveguide grating antennas for OPAs**

| Technology | Fabrication process | Feature size (nm) | Beam width (°) | $\frac{\partial \theta}{\partial \lambda}$ (°/nm) | Ref. |
|---|---|---|---|---|---|
| SOI | Single etch step | 80 | 0.10 | 0.13 | [81] |
| SOI | Single etch step | 100 | 0.081 | 0.17 | [82] |
| SOI | Single etch step | 80 | 0.2 | 0.14 | [97] |
| Si-SiN | Two etch steps | 500 | 0.2 | 0.094 | [90] |
| Si-SiN | Two etch steps | 315 | 0.075 | 0.175 | [41] |
| Si-SiN | Two etch steps | 90 | 0.2 | 0.074 | [98] |
| LN | Two etch steps | 100 | 0.6 | 0.08 | [45] |
| SOI | Two etch steps | 110 | 0.13 | 0.237 | This work |

**Funding.** National Natural Science Foundation of China (62205054, 62375051); National Key Research and Development Program of China (2019YFB2205303); Natural Science Foundation of Jiangsu Province (BK20210207); Fundamental Research Funds for the Central Universities.

**Disclosures.** The authors declare no conflicts of interest.

**Data availability.** Data underlying the results presented in this paper are not publicly available at this time but may be obtained from the authors upon reasonable request.